\documentclass[aps,pra,reprint,nofootinbib,twocolumn,superscriptaddress,showpacs,showkeys,longbibliography]{revtex4-1}
\usepackage{eurosym}
\usepackage{amsmath,amssymb,amstext}
\usepackage{ulem} 
\usepackage[usenames,dvipsnames]{color}
\usepackage{graphicx}
\usepackage{braket}
\usepackage{natbib}
\usepackage{comment}
\usepackage{dcolumn}
\usepackage{wasysym}

\usepackage[colorlinks,bookmarks=false,citecolor=blue,linkcolor=red,urlcolor=blue]{hyperref}

\begin{document}
%%%%%%%%%%%%%%%%%%%%%%%%%%%%%%%%%%%%%%%%%%%
\title{Drag force of a exciton-polariton condensate under non-resonant pumping}
\author{Pei-Song He}
\affiliation{Department of Physics, Beijing Technology and Business University, Beijing 100048, China}
\author{Zhaoxin Liang}
\email{The corresponding author: zhxliang@gmail.com}
\affiliation{Department of Physics, Zhejiang Normal University, Jinhua, 321004, China}
\date{\today}
\begin{abstract}
Exciton-polariton condensate in semiconductor microcavities constitute a novel kind of non-equilibrium superfluid. In a recent experiment [P. Stepanov, {\it et. al.,} Nat. Commun. {\bf 10}, 1038 (2019)],  the dispersion relation 
of collective excitations in a polariton condensate under the resonant pumping has been investigated with the emphasis on the role of reservoir of long-lived excitons in determining the superfluidity. Inspired by such an experimental advance, we study the superfluidity of a exciton-polaritonn condensate under non-resonant pumping by calculating the drag force exerted on a classical impurity moving in a polariton condensate.  
For a non-resonant pumped polariton condensate prepared in the gapped phase, due to the reservoir's modes, the drag force can be large when the velocity of the impurity is small.  Besides, as the velocity increases, the drag force can decrease.  For not very large velocity, the drag force is enhanced if the condensate is tuned to be more dissipative.  When the condensate is close to the transition point between the gapped phase and the gapless one, the drag force is similar to that of the equilibrium superfluid. Our present work reveals the effects of the reservoir's modes on the superfluidity properties of a polariton condensate with the non-resonant pumping.

\end{abstract}
\maketitle

\section{Introduction}

At present, there exist significant interests and ongoing efforts of studying the non-equilibrium superfluidity~\cite{SupFluidRev1,SupFluidRev2,SupFluidRevKeeling2007,SupFluidRevKeeling2011,Shelykh2009} of a exciton-polariton condensate, which conceptually goes beyond the equilibrium superfluidity such as liquid helium and Bose-Einstein condensates (BEC) of ultracold atoms~\cite{AtomicBECRev}. The motivation behind these interests is twofold. On the one hand, a polariton condensate is intrinsically non-equilibrium with coherent and dissipative dynamics occurring on an equal footing.  Similar to the ultracold atomic BEC,  the typical aspects of superfluid behavior of a polariton condensate such as quantized vortices~\cite{Lagoudakis2008,Nardin2011,Sanvitto2010}, suppression of scattering off disorder~\cite{Amo20091,Amo20092}, and the Bogoliubov's excitations~\cite{Utsunomiya2008,Kohnle2011}  have been observed experimentally. On the other hand, going beyond the equilibrium superfluid, the intrinsically non-equilibrium nature of a polariton condensate captured by the dissipation has induced the Goldstone mode~\cite{Wouters2007,Littlewood2006,Byrnes2012,Xu2017} of a spatially homogeneous system to be diffusive and made ghost branches~\cite{Pieczarka2015} of Bogoliubov excitations to be observable. These remarkable properties of the non-equilibrium superfluid prompt questions about superfluidity definition of superfluidity and characteristic observables~\cite{Janot2013,Keeling2011,Van2014,Gladilin2016,Juggins2018}. Further work based on spatially homogeneous fluids  has formulated generalized nonequilibrium versions of the Landau critical velocity~\cite{Keeling2011}.  Up to now, a polariton condensate has emerged as a novel kind of quantum superfluids characterized by its non-equilibrium characters~\cite{SupFluidRev1,SupFluidRev2}.

In the previous work, the intrinsically non-equilibrium nature of a polarion condensate is usually referred to as the intrinsic dissipation.  Very recently, a group from Institu N\'eel in France~\cite{Stepanov2019} has revealed the role of the another kind of intrinsically non-equilibrium nature of a polariton condensate due to the reservoir of long-lived excitons~\cite{Sarkar2010,Walker2017}. The nontrivial results in Ref.~\cite{Stepanov2019} consist of a speed of sound being apparently twice too low, which can not be explained upon considering the polariton condensate alone. In a combined theoretical and experimental analysis, Ref.~\cite{Stepanov2019} demonstrated that the presence of an excitonic reservoir alongside the polariton condensate has a dramatic influence on the characteristics of the quantum fluid. Motived by Ref. \cite{Stepanov2019}, an immediate theoretical work~\cite{Amelio2020} has investigated superfluidity in a nonequilibrium polariton fluid injected by a  coherent pump and flowing against a defect located outside the pump spot. The role of a reservoir in inducing nonstationary behaviors with moving phase singularities is  highlighted. 

However, the theoretical treatment of Ref. \cite{Amelio2020} is limited within the case of a polariton condensate with the resonant pumping, where the long-range
coherence is directly imprinted by the laser, and not the result of a condensation mechanism. Meanwhile, the non-resonant pumping
allows one to create a condensate with spontaneously chosen phase profile. Considered that the phase of a condensate plays a key role in determining the superfluidity properties, a timely question along the research line of Ref. \cite{Amelio2020}  arises as what is role of a reservoir in determining the superfluidity of a polariton condensate under the non-resonant pumping? 

In this paper, we are motivated to study the superfluidity of a polariton condensate under a non-resonant pumping by calculating the drag force on a moving impurity in a dynamically stable polariton condensate in the gapped phase.  The velocity of the impurity is set to be constant.  We have calculated the pressure force due to density gradient of the condensate around the impurity, which are outcomes of excitations created by the moving impurity.  The frictions due to scattering between the impurity and the reservoir polaritons are not considered here for simplicity. We find that when the velocity of the impurity is small, the impurity will experience a large drag force due to the slow diffusion of the reservoir's modes at small momentum.  We also find that the drag force can be decreased when the velocity of the impurity is increased, this is also due to the effect of the diffusive mode.  Besides, we find that the drag force is enhanced as the condensate is tuned to be more dissipative.

The emphasis and value of the present work are to study the role of the non-resonant reservoir on the non-equilibrium superfluidity by analyzing the combined effects of both condensate modes and the reservoir modes on the drag force. In this context,  differently from the calculation of the drag force for a non-equilibrium condensate in Ref.~\cite{Pinsker2017} based on the complex Gross–Pitaevskii (GP) theory, our model fully includes the role of the non-resonant reservoir by solving driven-dissipative GP equation coupled to a rate equation; differently from Ref.~\cite{WountersPRL2010,Amelio20201} of calculating the drag force of non-resonantly pumped polariton condensate prepared in the gapless region of the Bogoliubov modes, we study the drag force of non-resonantly pumped polariton condensate in the gapped region of the Bogoliubov modes. We believe the calculations in this work together with Ref.~\cite{Stepanov2019,Amelio2020} can provide a complete description of the role of a reservoir in inducing nonequilibrium superfluid behaviors for a polariton condensate.

This paper is organized as follows. In Sec. \ref{sec_model},  we briefly describe the model system based on driven-dissipative Gross-Pitaevskii equation. In Sec. \ref{Sec_Drag}, the definition and analytical expressions of the drag force are given by solving the Bogoliubov–de Gennes equations.  In Sec. \ref{sec_results}, we give the detailed results and analysis of the drag force in the different phases.  In Sec. \ref{sec_conclusions}, we give the conclusions of this work and some discussions.

\section{Model and Formalism}
\label{sec_model}

Our goal is to calculate the drag force exerted on a classical impurity moving in a  polariton condensate, in particular, focus on the reservoir's effects on the non-equilibrium superfluidity. To this end, we consider a exciton-polariton BEC created under non-resonant pumping with a point-like impurity moving in the polarition condensate and are interested in the drag force exerted on the moving impurity. The order parameter for the condensate is described by a one-component time-dependent wave function $\psi\left({\bf r},t\right)$; the reservoir on the relevant time scales can be modeled by a scalar density denoted by $n_R\left({\bf r},t\right)$.  

At the mean-field level, both the static and dynamical properties of the polariton condensate can be described by the driven-dissipative Gross-Pitaevskii equation (GPE)~\cite{Wouters2007,Xu2017}, i.e.,
\begin{eqnarray}
 	i\hbar\frac{\partial\psi}{\partial t} = \Big[-\frac{\hbar^2}{2m_{\text {LP}}}\nabla^2 &+& g_C |\psi|^2 + g_Rn_R \nonumber\\
			& +& \frac{i\hbar}{2}\left(Rn_R-\gamma_C\right) + V_{\text{im}} \Big]\psi.
\label{eqn_gpe_1}
\end{eqnarray}
Here, $m_{\text {LP}}$ is the mass of the polaritons, $g_C$ is the interaction constant between coherent condensate polaritons and $g_R$ characterizes the interaction between the condensate and the incoherent reservoir of $n_{\text{R}}$.  $R$ stands for the stimulated scattering rate, which replenishes the condensate from the reservior polaritons and $\gamma_C$ is the rate of loss of condensate polaritons with the main effect of leaking out of the photons. In this work, 
we focus on a point-like impurity moving in the polarition condensate. We remark that we do not consider the effect arise from interactions between the impurity and the density of reservoir polaritons $n_R(\mathbf{r})$ for simplicity.
By including the impurity potential and based on a mean-field description of the condensate in terms of the GP Eq. (\ref{eqn_gpe_1}) for the macroscopic wave function $\psi(\mathbf{r})$ including the impurity potential as follows 
\begin{equation}
	V_{\text{im}}(\mathbf{r}) = g_i\delta(\mathbf{r}-\mathbf{v}t),\label{Disorder}
\end{equation}
with $g_i$ being the $s$-wave scattering length between the impurity and polariton and $\mathbf{v}$ labelling the velocity of the impurity.

We consider Eq. (\ref{eqn_gpe_1}) is coupled to a scalar incoherent reservoir as mentioned earlier, described by a rate equation~\cite{Wouters2007,Xu2017}, i.e.,
\begin{equation}
 	\frac{\partial n_R}{\partial t} = P - \left(\gamma_R+R|\psi|^2\right)n_R + D\nabla^2 n_R,
\label{eqn_gpe_2}
\end{equation}
Here, $P$ is the rate of an off-resonant continuous-wave (CW) pumping, $\gamma_{\text{R}}^{-1}$ describes the lifetime of reservoir polaritons, $D$ reads the diffusive constant of the reservoir
and $R$ is the stimulated scattering rate of reservoir polaritons into the condensate.

In this work, we limit ourselves to the case that the impurity potential of Eq. (\ref{eqn_gpe_2}) can be safely treated as the weak perturbation on the polariton system, which means that the modifications on the properties of the macroscopic wave function $\psi$ and the excitations over it can be safely neglected~\cite{Astrakharchik2004}.  In this way, the impurity moves in the same stationary state as that without the impurity potential.  

\section{Drag force}\label{Sec_Drag}
We focus on investigating the role of a reservoir in determining the superfluidity of a non-resonantly pumped polariton condensate.  At the heart of our solution of the non-equilibrium superfluidity is
to calculate the drag force exerted on a classical impurity moving in a  polariton condensate with the emphasis on the effects of reservoir modes on the drag force.

As the first step, we need to determine the phase diagram of the stationary states described by Eqs. (\ref{eqn_gpe_1}) and (\ref{eqn_gpe_2}). As mentioned before, the moving potential of Eq. (\ref{Disorder}) 
has been treated as a perturbation and can be safely ignored in determined the stationary states. In such,  when $P$ is larger than the threshold $P_{\text{th}} = \gamma_C\gamma_R / R$, due to vanishing of net gain imposed by stationarity, the reservoir density is clamped to $n_R^0 = \gamma_C/R$, while the condensate density has $n^0_C = |\psi_0|^2 =  (P-P_{\text{th}}) / \gamma_C$~\cite{Wouters2007}.  The oscillation frequency of the macroscopic wave function is $\mu_0 = g_C n^0_C + g_R n_R^0$.

Next, we need the knowledge of collective excitations induced by the moving potential of Eq. (\ref{Disorder}) . Within the mean-field framework, we follow the standard
procedures in Refs. \cite{Wouters2007,Xu2017} and start from the standard decomposition
of the wave function ($\psi$, $n_R$) into the steady-state solution ($\psi_0$, $n^0_R$) and a small fluctuating term ($\delta n_C$, $\delta\theta_C$, $\delta n_{R}$), i. e.,
\begin{eqnarray}
 	\psi &=& e^{-i\mu_0 t/\hbar}\sqrt{n^0_C + \delta n_C} e^{i\delta\theta_C}, \nonumber\\
     n_R &=& n_R^0 + \delta n_R.
\label{eqn_excitation}
\end{eqnarray}
Substituting (\ref{eqn_excitation}) into eqns. (\ref{eqn_gpe_1}) and (\ref{eqn_gpe_2}) and retaining only first-order terms of fluctuation, we can obtain the Bogoliubov–de Gennes (BdG) equations for $\mathbf{V}(\mathbf{k},t) \equiv [\delta n_C(\mathbf{k},t), \delta\theta_C(\mathbf{k},t), \delta n_{R}(\mathbf{k},t)]^T $ as follows \cite{Astrakharchik2004,Carusotto2004,CarusottoBEC2006,Wouters2010}
\begin{equation}
 	\frac{\partial \mathbf{V}(\mathbf{k},t)}{\partial t} = \mathbf{M}(\mathbf{k})\mathbf{V}(\mathbf{k},t) + \mathbf{B}(\mathbf{k},t),
\label{eqn_diff}
\end{equation}
with
\begin{equation}
     \mathbf{M}(\mathbf{k}) =  \begin{pmatrix}
		0&\frac{\hbar n^0_C k^2}{m_{\text{LP}}}&Rn^0_C\\
		-\frac{\hbar k^2}{4n^0_C m_{\text{LP}}}-\frac{g_C}{\hbar}&0&-\frac{g_R}{\hbar}\\
		-Rn_R^0&0&-Dk^2 - \gamma_R - Rn^0_C
	  \end{pmatrix}, 
\end{equation}
and
\begin{equation}
    \mathbf{B}(\mathbf{k},t) = - \frac{g_i}{2\pi\hbar} e^{-i\mathbf{k}\cdot\mathbf{v}t}\begin{pmatrix}0\\1\\0\end{pmatrix}.
\end{equation}

Finally,  the force exerted on the polaritons in the coherent condensate state due to the delta potential imposed by the moving impurity can be expressed as~\cite{He2014,Astrakharchik2004,Wouters2007,Pinsker2017BEC}
\begin{eqnarray}
 	\mathbf{F}(t) &=& -\int d\mathbf{r}|\psi(\mathbf{r},t)|^2\nabla[g_i\delta(\mathbf{r}-\mathbf{v}t)] \nonumber\\
 	&=& g_i \left[\nabla|\psi(\mathbf{r},t)|^2\right]\big|_{\mathbf{r}=\mathbf{v}t} \nonumber\\
 	&=& g_i \int \frac{d\mathbf{k}}{2\pi} i\mathbf{k}\delta n_C(\mathbf{k},t)e^{i\mathbf{k}\cdot\mathbf{v}t}.
\label{eqn_dragforce_formula}
\end{eqnarray}
The force on the impurity is just the minus of the above result of Eq. (\ref{eqn_dragforce_formula}), showing that the drag force originates from the scattering between the impurity and the induced density fluctuation.
We point out that the force induced by scattering between the impurity and the incoherent reservoir has beyond the scope of this paper.  The latter is proportional to the density of the incoherent reservoir of $n_R$; while the former is related to the condensate fluctuation of $\delta n_C(\mathbf{k},t)$.  From the second line in Eq. (\ref{eqn_dragforce_formula}), we remark that the drag force we defined is just the pressure force due to density difference around the impurity, that is, the force is due to density accumulation after excited by the moving impurity. 

In order to obtain the analytical expression of the drag force of Eq. (\ref{eqn_dragforce_formula}), we need the knowledge of the $\delta n_C(\mathbf{k},t)$ by exactly solving the differential equation (\ref{eqn_diff}) as follows
\begin{equation} 
 	\mathbf{V}(\mathbf{k},t) = e^{(t-t_0)\mathbf{M}(\mathbf{k})}\mathbf{V}(\mathbf{k},t_0) + \int^{t}_{t_0}e^{(t-s)\mathbf{M}(\mathbf{k})}\mathbf{B}(\mathbf{k},s)ds.
\end{equation}
Note that only the second term on the right hand of the above equation is proportional to the strength of the impurity potential.
We will consider the steady state with the impurity moving with a constant velocity in the condensate.  This can be obtained by taking $t_0 = -\infty$, which means to turn on the impurity potential adiabatically.
Then the impurity-induced excitations over the coherent polariton condensate read~\cite{Harris2001,He2014}
\begin{eqnarray}
 	\delta n_C(\mathbf{k},t) &=& \int^{t}_{-\infty} e^{(t-s)\mathbf{M}(\mathbf{k})}\mathbf{B}(\mathbf{k},s)ds \nonumber\\
 	&=& -\frac{g_i n^0_C k^2}{2\pi m_{LP}} e^{-i\mathbf{k}\cdot\mathbf{v}t}\int^{\infty}_{0}ds  e^{i\mathbf{k}\cdot\mathbf{v}s}\Big[(\lambda_2^2-\lambda_3^2)e^{\lambda_1 s} \nonumber\\
	&&+ (\lambda_3^2-\lambda_1^2)e^{\lambda_2 s} + (\lambda_1^2-\lambda_2^2)e^{\lambda_3 s}\Big]\nonumber\\
	&&\Big/\Big[(\lambda_1-\lambda_2)(\lambda_2-\lambda_3)(\lambda_3-\lambda_1)\Big],
\label{eqn_density_var}
\end{eqnarray}
where $\lambda_j(\mathbf{k})$ ($j=1,2,3$) are the three eigenvalues of the matrix $\mathbf{M}(\mathbf{k})$, i.e. $\det(\mathbf{M}-\lambda \mathbf{I})=0$. In more details, the $\lambda_j(\mathbf{k})$ ($j=1,2,3$) are the 
three roots of the following equation, reading
\begin{equation}
 	\lambda^3 + b\lambda^2 + c\lambda +d = 0,
\label{eqn_eigenvalue}
\end{equation}
with
$b = Dk^2 +(1+\alpha)\gamma_R, 	c = \varepsilon_{\mathbf{k}}^2/\hbar^2 + \alpha\gamma_R\gamma_C$, 	$d = b\varepsilon_{\mathbf{k}}^2/\hbar^2 - \alpha g_R\gamma_R\gamma_C k^2/(Rm_{LP})$, 	$\varepsilon_{\mathbf{k}} = \sqrt{\frac{\hbar^2 k^2}{2m_{LP}}\big(\frac{\hbar^2 k^2}{2m_{LP}}+2g_Cn^0_C\big)}$,  and	$\alpha = P/P_{\text{th}}-1$. We also point out that the corresponding dispersion relations of the excitations can be expressed in terms of  $\lambda_j(\mathbf{k})$ as 
\begin{equation}
	\omega_{j}(\mathbf{k}) = i \lambda_j(\mathbf{k}),\ \ \ j = 1, 2, 3.
\label{eqn_omega_lambda}
\end{equation}

Before analyzing the drag force, it is essential to establish that the homogeneous background itself
is stable with respect to weak perturbations. The condition for the condensate to be stable against perturbtions is that the excitations do not increase with time.  It requires that all of the $\lambda$'s in eqn. (\ref{eqn_density_var}) have zero or negative real parts.  In terms of dispersion relation $\omega$, it means that the imaginary part of $\omega$ should all be zero or negative for any momentum.

According to the Routh-Hurwitz stability criterion, the sufficient and necessary condition for all of roots of the cubit equation with real coefficients, $a_3 z^3 + a_2 z^2 + a_1 z + a_0 = 0$, to be in the left half plane is that: $a_i>0$, $i = 0,1,2,3$, and at the same time $a_1 a_2 > a_0 a_3$.  In this case, the system is stable against perturbations.
When applied to eqn. (\ref{eqn_eigenvalue}), we find that when $d>0$, there is always $bc>d$.  So the stability condition of the condensate against perturbations is that $d>0$, that is 
$P>P_{\text{th}} = P_{\text{th}}\cdot(\gamma_C/g_C)/(\gamma_R/g_R)$. In what follows, we restrict our consideration to the
dynamics of a classical impurity propagating on a modulationally
stable condensate background. Therefore we make
sure that the parameters of the system always satisfy the above 
condition.

\subsection{Excitations in the gapped region}

As a consequence of dissipation, the matrix of $M$ in Eq. (\ref{eqn_density_var}) is non-Hermitian, and Eqs. (\ref{eqn_eigenvalue}) and (\ref{eqn_omega_lambda}) yields three complex dispersion
branches where the imaginary part represents the damping spectrum. We denote $A = b^2 - 3c, B = bc-9d, C = c^2-3bd$, and $\Delta =  B^2 - 4AC$.  It can be obtained that when $\Delta\le0$, Equation (\ref{eqn_eigenvalue}) has three real solutions, while for $\Delta > 0$ the equation has one real solutions and two conjugate complex ones with both the real and imaginary parts finite.  The corresponding dispersion relation are then given by $\omega = i\lambda$.  Below, we first briefly review some important features of the Bogoliubov excitation modes.

\textit{ {\it Gappless} region:}   When $\gamma_R/\gamma_C > 4\alpha/(1+\alpha)^2$, the dispersion relations of three modes in Eqs. (\ref{eqn_eigenvalue}) and (\ref{eqn_omega_lambda}) are purely imaginary if the momentum is small, while for large enough momentum, two of the three modes becomes ones with finite real and imaginary parts.  This is referred as to the {\it gappless} region firstly discussed in Ref.~\cite{Wouters2007}, where the Bogoliubov mode becomes to be diffusive. Furthermore, the superfluidity of a polariton condensate based on the drag force in the {\it gappless} region has been investigated in Ref.~\cite{Wouters2010,Amelio2020} with emphasis on the effects of the intrinsic dissipation due to the non-equilibrium nature on the superfluidity.

\begin{figure}[t]
\center
\includegraphics[width=0.45\textwidth]{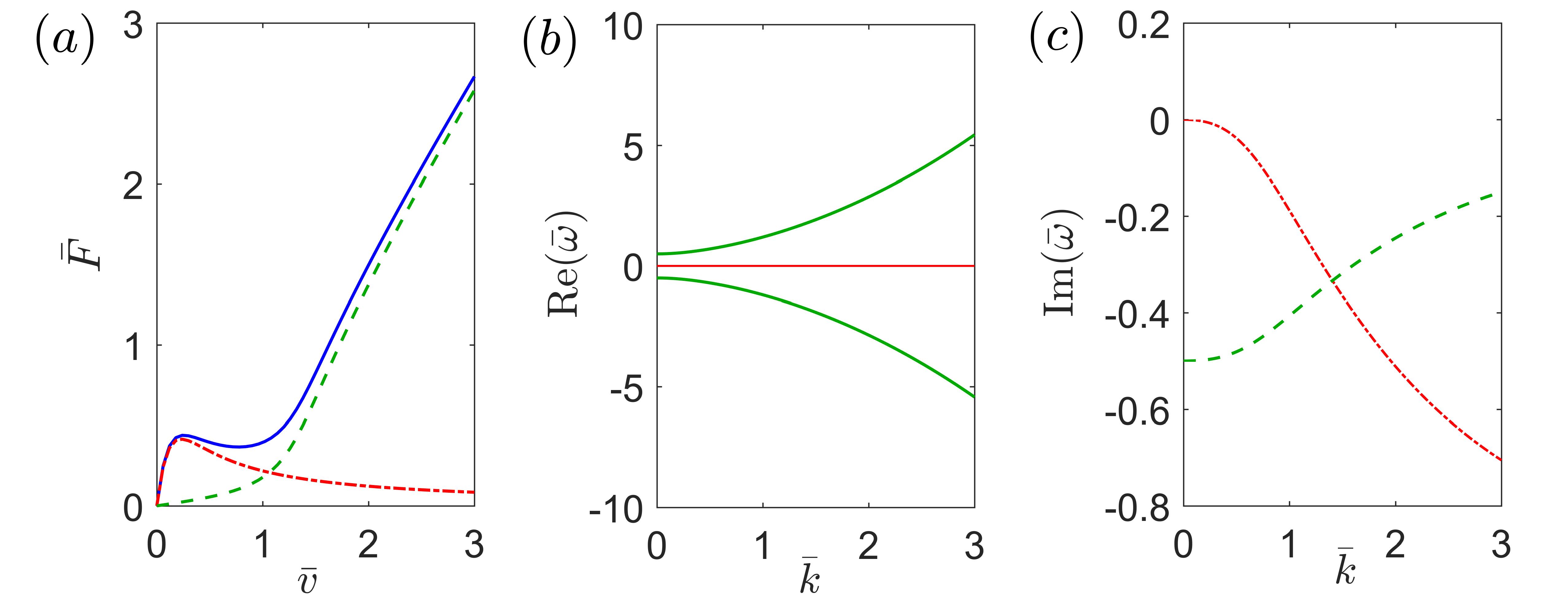}
\caption{(a) Drag force of Eq. (\ref{DragForce}) exerted on a classical impurity moving in a polariton condensate. The green dashed and red dash-dotted curves are represented by the drag forces due to the condensate's excitations and the reservoir's modes. The solid blue curve is drag force induced by both  the condensate's excitations and the reservoir's modes together. (b) and (c) are the real and imaginary parts of the Bogolubov's modes in Eq. (\ref{eqn_eigenvalue}). The parameters are given as $D_0 = 0.0005, \alpha = 1.0, \bar{\gamma}_C = 1.0$, and  $\bar{g}_R/(1+\alpha) : \bar{\gamma}_R/\bar{\gamma}_C : 4\alpha/(1+\alpha)^2 =0.475 :0.5:1.0$.  
}
\label{fig01}
\end{figure}

\textit{{\it Gapped} region:}  When $\gamma_R/\gamma_C < 4\alpha/(1+\alpha)^2$, there is always $\Delta>0$ for any momentum $\mathbf{k}$.  There are two  conjugate dissipative gapped modes and one diffusive modes with a purely imaginary dispersion relation, such as that shown in Fig. \ref{fig01} (b) and (c).  We call this {\it gapped} region~\cite{Byrnes2012}, in which the properties of superfluidity based on the drag force have not been discussed yet.  In what follows, we are interested in the superfluidity of a polariton condensate in the gapped region by calculating the drag force on a moving impurity. We remark that our work together with Ref.~\cite{Wouters2010,Amelio20201} give the complete description of superfluidity of a non-resonant pumped polartion condensate in the full parameter regions. 

\subsection{Drag force in gapped region}

Since all of the $\mathrm{Im}\omega$'s are negative, we can obtain from Eqs. (\ref{eqn_dragforce_formula}) and (\ref{eqn_density_var}) that the drag force has the form
\begin{widetext}
\begin{eqnarray}
 	\mathbf{F} 
     &=&
- i \hat{x}\frac{g_i^2 n_C^0 v}{\pi^2} \int^{\infty}_0 dk_x \int^{\infty}_0 dk_y \frac{k_x^2 k^2}{(\omega_1-\omega_2)(\omega_2-\omega_3)(\omega_3-\omega_1)} \left[\frac{\omega_2^2 - \omega_3^2}{\omega_1^2 - (k_x v)^2} + \frac{\omega_3^2 - \omega_1^2}{\omega_2^2 - (k_x v)^2} + \frac{\omega_1^2 - \omega_2^2}{\omega_3^2 - (k_x v)^2}\right].
\label{eqn_force_omega}
\end{eqnarray}
\end{widetext}
Equation (\ref{eqn_force_omega}) is one of the main results obtained in this paper, from which we immediately notice that all of the contributions from the three excitations to the drag force are positive.

From the expression of the drag force in Eq. (\ref{eqn_force_omega}), we can see that since the imaginary parts of both the dissipative modes and the diffusive modes are finite, there is always drag force for any finite velocity of the impurity, whether the Landau's criterion for a equilibrium superfluid $\mathrm{Re}\omega(\mathbf{k}) - \mathbf{k}\cdot\mathbf{v} = 0$ is satisfied or not.  So the superfluid critical velocity is zero for our situation.  Both the dissipative gapped condensate's modes and the diffusive reservoir's modes give nonzero contributions to the drag force for any finite velocity.  
%For the gapped modes, due to 
%For the gapped modes, it is because of the dissipative nature of these modes that they can be excited even when the Landau's criterion is not satisfied.  For the diffusive modes, the diffusive properties of these modes make that any finite energy can creat them and lead to loss of the momentum and energy.
%For the gapped modes, since it requires finite energy to excite these modes, there is energy dissipation when the impurity moves.  For the diffusive modes with zero $\mathrm{Re}\omega$ but finite momenta, there is back force on the impurity when the modes are excited and diffuse.  So to maintain a constant velocity of the impurity, there should be an external force exerted on the impurity to balance this back force.  This is just the considered drag force contributed by the diffusive mode.

The drag force in Eq. (\ref{eqn_force_omega}) can be further rewritten with the help of Eq. (\ref{eqn_eigenvalue})
\begin{eqnarray}
 	\mathbf{F} &=&  \frac{g_i^2 n^0_C \mathbf{v}}{\pi^2 m_{LP}}\int^{\infty}_0 dk_x \int^{\infty}_{0} d k_y \nonumber\\
				   &&\times  \frac{k_x^2 k^2(bc-d)}{(k_x v)^2[(k_x v)^2-c]^2 + b^2[(k_x v)^2-d/b]^2}.
\end{eqnarray}
This expression is more convenient to calculate the drag force numerically, compared with the expression in Eq. (\ref{eqn_force_omega}). We further write the dimensionless drag force  $\bar{\mathbf{F}} = \mathbf{F}/(\hbar c_s n^0_C)$ as
\begin{eqnarray}
 	\bar{\mathbf{F}} &=&  \frac{\bar{g_i}^2 \bar{\mathbf{v}}}{\pi^2}\int^{\infty}_0 d\bar{k}_x \int^{\infty}_{0} d \bar{k}_y \nonumber\\
				   &&\times  \frac{\bar{k}_x^2 \bar{k}^2(\bar{b}\bar{c}-\bar{d})}{(\bar{k}_x \bar{v})^2[(\bar{k}_x \bar{v})^2-\bar{c}]^2 + \bar{b}^2[(\bar{k}_x \bar{v})^2-\bar{d}/\bar{b}]^2}.\label{DragForce}
\end{eqnarray}
In Eq. (\ref{DragForce}), all parameters are rewritten in the dimensionless form using healing length $r_h = \hbar/(m_{LP}c_s)$ and time $\tau_0 = r_h/c_s$, where $c_s = (g_Cn^0_C/m_{LP})^{1/2}$ is a local sound velocity in the condensate.
The dimensionless parameters are $\bar{g}_R = g_R / g_C, \ \ \bar{g}_i = g_i / (\hbar^2/m_{LP}), \bar{\gamma}_R = \gamma_R\tau_0, \ \ \bar{\gamma}_C  = \gamma_C\tau_0$, $\bar{P}_{th} = \bar{\gamma}_R\bar{\gamma}_C / \bar{R}$, $\bar{P} = (1+\alpha)\bar{P}_{th}$, and $\bar{D} = D / (\hbar/m_{LP})$.  Besides, $\bar{n}^0_R = n^0_R / n_C^0 = \bar{\gamma}_C / \bar{R}$.
The dimensionless dispersion relations are solved from $\bar{\lambda}^3 + \bar{b}\bar{\lambda}^2 + \bar{c}\bar{\lambda} + \bar{d} = 0$, where $\bar{b} = \bar{D}\bar{k}^2 +(1+\alpha)\bar{\gamma}_R, 	\bar{c} = \bar{\varepsilon}_{\mathbf{k}}^2 + \alpha\bar{\gamma}_R\bar{\gamma}_C$, $\bar{d} = \bar{b}\bar{\varepsilon}_{\mathbf{k}}^2 - \bar{g}_R\bar{\gamma}_C\bar{k}^2$.  
So $\bar{\mathbf{F}}$ depends on $\bar{D}, \bar{\gamma}_R, \bar{\gamma}_C, \bar{g}_R, \alpha, \bar{g}_i, \bar{v}$. 
We study in this paper the gapped region that is dynamically stable, which means that
$\bar{g}_R/(1+\alpha) < \bar{\gamma}_R/\bar{\gamma}_C < 4\alpha/(1+\alpha)^2$.

\section{Results}
\label{sec_results}

In the previous section, we have developed the intuitive physical picture and predicted features in the drag force of a polariton condensate under the non-resonant pumping.
Below we study the drag force  exerted on a classical impurity moving in a polariton condensate with emphasis on the effects of the non-resonant pumping by self-consistently solving Eqs. (\ref{eqn_eigenvalue}), (\ref{eqn_omega_lambda}), and (\ref{DragForce}) numerically. In this end,  our strategy is to (i) calculate the drag force induced by the condensate's modes and the non-resonant reservoir's mode separately; (ii) analyze how the non-resonant reservoir's modes affect the behavior of drag force; (iii) add the drag force induced by the condensate's modes and the non-resonant reservoir's mode together and then investigate superfluidity of a polariton condensate highlighted by effects of the non-resonant pumping. 

We begin to calculate the drag force according to Eq. (\ref{DragForce}) for a polariton condensate prepared in the {\it gapped} regime of the Bogoliubov's modes of Eq. (\ref{eqn_omega_lambda}) and (\ref{DragForce}) as mentioned in Sec. \ref{Sec_Drag} A. For the convenience of the later analysis, we denote the dispersion relation of the dissipative gapped modes of a condensate as $\bar{\omega}_{1, 2}$ with $\mathrm{Re}(\bar{\omega}_1) = - \mathrm{Re}(\bar{\omega}_2) >0$, $\mathrm{Im}(\bar{\omega}_1) = \mathrm{Im}(\bar{\omega}_2) <0$, and that of the non-resonant reservoir's mode as $\bar{\omega}_3$ with $\mathrm{Re}(\bar{\omega}_3) = 0$ and $\mathrm{Im}(\bar{\omega}_3) < 0$. In particular,  we note that $\mathrm{Im}\bar{\omega}_3(\bar{k}) \approx -i\bar{\Gamma}\bar{k}^2$ for small momentum $\bar{k}$ with the effective diffusive constant $\bar{\Gamma} = \frac{1+\alpha}{\alpha\bar{\gamma}_R}(\frac{\bar{\gamma}_R}{\bar{\gamma}_C} - \frac{\bar{g}_R}{1+\alpha})$ \cite{Xu2017}.

We are ready to study how the existence of the non-resonant pumping affects the drag force with the help of Eq. (\ref{DragForce}). For this purpose,  we devise three scenarios: first, we choose the proper parameters of Bogoliubov's modes and find the drag force in the {\it gapped} region; then we vary the parameter of $\bar{g}_R$ with all of the other parameters of the system fixed and study how the non-resonant reservoir affects the drag force by the $\bar{g}_R$; Finally,  we study how the band gap labelled by $ \alpha\bar{\gamma}_R\bar{\gamma}_C$ affects behavior of the drag force. 

\begin{figure}[t]
\center
\includegraphics[width=0.45\textwidth]{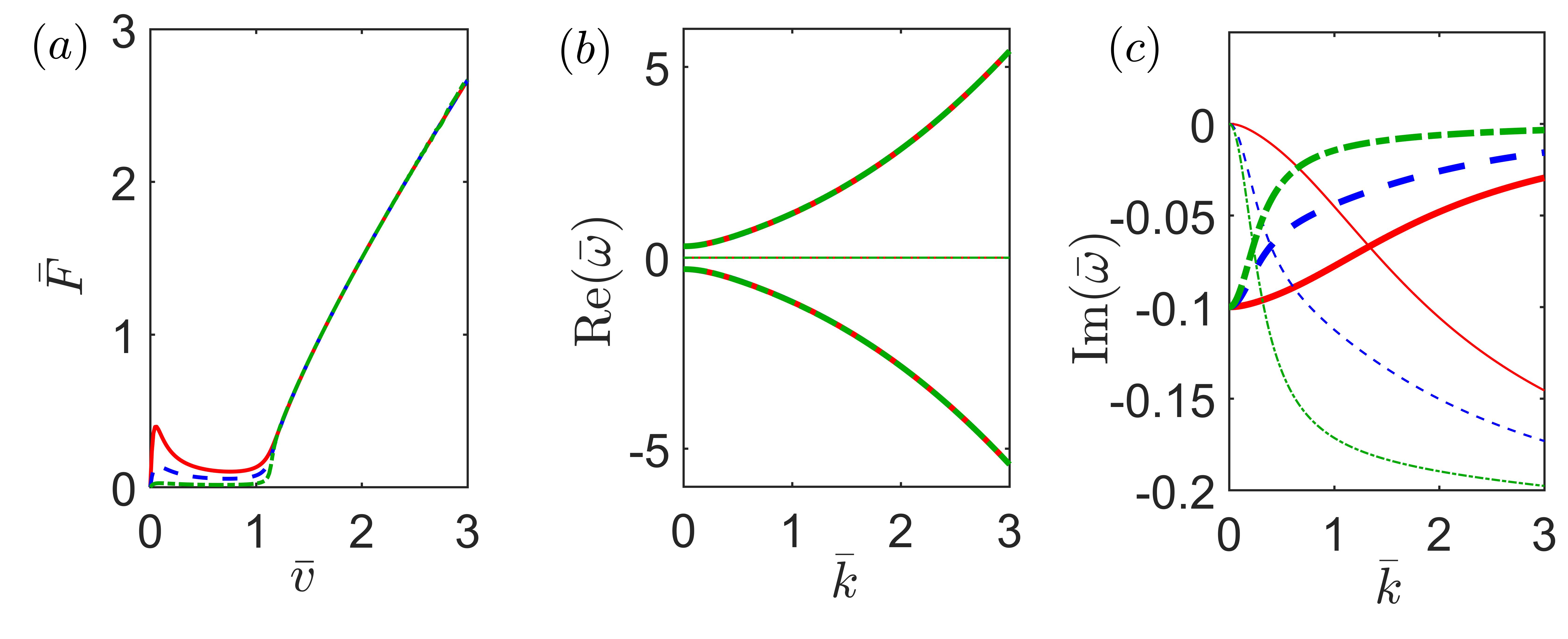}
\vspace{-6pt}
\caption{ (a) The drag force in gapped region for fixed $\bar{\gamma}_C = 1.0$,  $D_0 = 0.0005$ and $\alpha = 1.0$ with $\bar{\gamma}_R/\bar{\gamma}_C : 4\alpha/(1+\alpha)^2 = 0.1$ with $\bar{g}_R/(1+\alpha)$ : $\bar{\gamma}_R/\bar{\gamma}_C = 0.1$ (\text{green dash-dotted}), 0.5 (\text{blue dashed}), 0.95 (\text{red solid}).  
As $|\mathrm{Im}(\bar{\omega})|$ of the diffusive modes decreases, the contributed diffusive force leads to an enhanced total drag force for $\bar{v}$ small.  (b) and (c) are represented by the real and imaginary parts of the Bogolubov's modes in Eq. (\ref{eqn_eigenvalue}). %The thick and thin lines represents the gapped modes and the purely imaginary modes, respectively.
}
\label{fig11}
\end{figure}

In the the first scenario where Bogoliubov's modes are located in the ${\it gapped}$ region (see Figs. \ref{fig01} (b) and (c)), the corresponding drag forces are plotted in Figs. \ref{fig01} (a).
As shown in Figs. \ref{fig01} (a),  we find that the drag force induced by the reservoir's modes increases in a steep way for small velocity of the impurity, and then goes smaller gradually for large velocity, in between a hump appears (see the red curve in Figs. \ref{fig01} (a)). This gives the main contributions to the total drag force when the velocity of the impurity is small (see solid blue curve in Figs. \ref{fig01} (a)).
Besides, the part of the drag force contributed by the codensate's gapped modes is nonzero when the velocity of the impurity is below the classical Landau's critical superfluid velocity (see the green curve in Figs. \ref{fig01} (a)), which is an effect of the dissipation of condensates.
When the velocity of the impurity is large, this part of drag force grows approximately linear in velocity, like that of the equilibrium condensates~\cite{Astrakharchik2004} and consists the main contributions to the total drag force.

In the second scenario,  we consider the situation with a variation of $\bar{g}_R$ while all of the other parameters of the system are kept to be fixed.  
Then a larger $\bar{g}_R$ means a smaller effective diffusion constant $\bar{\Gamma}$.
In Fig. \ref{fig11}, we consider the variation of the condensate from close to the dynamically instability point (the red solid line) to that deep in the stable phase (the green dash-dotted line).
%Physically, the parameters choosen in Fig. \ref{fig11} corresponds to variation of the condensate from close to the dynamically instability point to that deep in the stable phase.

In Figs. \ref{fig11} (a), we find that as $\bar{\Gamma}$ is reduced, the drag force at small $\bar{v}$ increases prominently.
Physically, a smaller $\bar{\Gamma}$ means a slower diffusion of the density accumulation around the impurity, which leads to a larger pressure force on the impurity.
This is one of the main features we found in the gapped phase. For $\bar{v}$ in the intermediate region, the drag force can be much smaller than the one for small $\bar{v}$.  This is an interesting phenomenon neither common in equilibrium quantum systems nor in nonequilibrium classical ones.
For $\bar{v}$ in the intermediate region, we also find that the drag force is enhanced as $\bar{g}_R$ increases.  
It is because the drag force contributed by the dissipative gapped modes goes larger as the amplitudes of the imaginary part of the dispersion relation, which characterize the dissipativeness of the condensate, are enhanced.
For $\bar{v}$ large, the drag force is approximately linear in velocity, similar to that of the equilibrium BEC.
%In Region III of the drag force as in Fig. \ref{fig11}, it is approximately linear in velocity, which is similar to that of the equilibrium BEC.  
%In this region, the velocity of the impurity is large enough to create modes satisfy $\mathrm{Re}\bar{\omega}_{1,2}(\bar{\mathbf{k}}) - \bar{\mathbf{k}}\cdot\bar{\mathbf{v}} = 0$.  Furthermore, $\mathrm{Im}\bar{\omega}_{1,2}(\bar{\mathbf{k}})$ is small for large $\bar{k}$.  Therefore, this situation is similar to that of the equilibrium BEC, in which the excitations satisfy the Landau's criterion, and the excitations are dissipativeless travelling waves.  This is consistent with the expression in eqn. (\ref{eqn_force_III}).
%On the other hand, since $|\mathrm{Im}(\bar{\omega}_3)|$ is large at large $\bar{k}$, the excitations created diffuse quickly, and the induced drag force is small.  
The total drag force is almost contributed by the dissipative gapped modes.
%, which resemble that of the equilibrium BEC.

%As a summary of this subsection, we find that the hump with a sharp slope at small velocity of the impurity in Region I is due to the slow diffusion of the diffusive mode with small momentum $\bar{k}$.  Besides, the size of the drag force in Region II is larger if the condensate is more dissipative.  The drag force in Region III is similar to that of the equilibrium BEC since the dissipation is small and approximately the Landau's criterion to excite collective modes applies.

\begin{figure}[t]
\center
\includegraphics[width=0.48\textwidth]{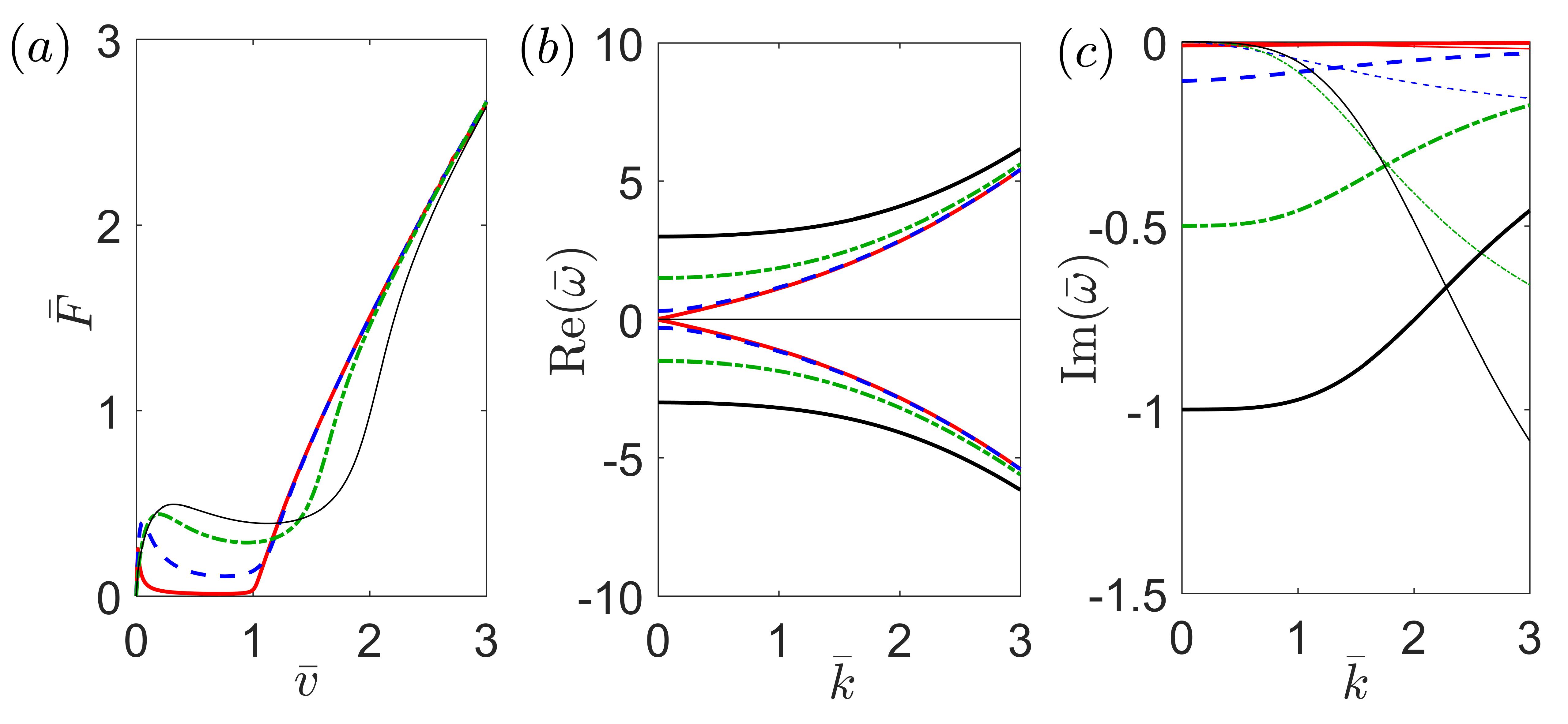}
\caption{(a) The drag force in gapped region for fixed $D_0 = 0.0005$, $\alpha = 1.0$ and $\bar{g}_R/(1+\alpha) : \bar{\gamma}_R/\bar{\gamma}_C : 4\alpha/(1+\alpha)^2 = 0.095:0.1: 1.0$ with $\bar{\gamma}_C = 0.1$ (\text{red solid}), $1.0 (\text{blue dashed})$, $5.0$ (\text{green dash-dotted}), $10.0$ (\text{thin black solid}).  (b) and (c)  are the real and imaginary parts of the Bogoliubov's modes respectively.
The dissipative force decreases as the $|\mathrm{Re}(\bar{\omega})|$ of the dissipative modes decreases, which leads to a suppressed total drag force for intermediate $\bar{v}$.
}
\label{fig21}
\end{figure}

%%%%%%%%%%%%%%%%%%%%%%%%%%%%%%%

In the third scenario,  we consider the variation of the drag force as the size of the band gap varies, which can be achieved by modifying the value of $ \alpha\bar{\gamma}_R\bar{\gamma}_C$. In Figs. \ref{fig21} (a), we fix the value of $\alpha$ and change that for both $\bar{\gamma}_R$ and $\bar{\gamma}_C$, but keep the ratio $\bar{g}_R/(1+\alpha) : \bar{\gamma}_R/\bar{\gamma}_C : 4\alpha/(1+\alpha)^2$ fixed.  
We find that as the condensate is tuned close to the transition point between the gapped phase and gapless one, both of the $|\mathrm{Im}\omega_{1,2}|$ and $|\mathrm{Im}\omega_3|$ are small, and the behavior of the drag force approaches that of the equilibrium BEC.  The condensate is lesser dissipative, and the drag force is smaller for $\bar{v}$ with intermediate value.  However, due to the small effective diffusive constant, there is still a large drag force for small $\bar{v}$.

Before making a conclusion, we want to remark the main results of the drag force of a non-resonant pumped polariton condensate in the gapless phase as studied in Ref.~\cite{Carusotto2004}. For a polariton prepared in the gapless region, the imaginary part of the reservoir modes is finite for all momenta, so the diffusion constant for small momenta is much larger than the one in the gapped phase.  It results that for slow motion of the impurity the drag force induced by the diffusive modes is more prominent in the gapped phase than that in the gapless phase. Besides, in the gapless phase, the real part of the dispersion relation of the dissipative modes is zero for small momenta.  From the Landau's criterion, for any small velocity of the impurity, there is creation of the dissipative modes, and it results in a finite drag force.  While in the gapped phase, for small velocity of the impurity, the Landau's criterion is not satisfied, and the drag force contributed by the dissipative modes are simply due to the dissipative nature of the modes.  In a consequence, the induced drag force by the dissipative modes is generally smaller in the gapped phase than that in the gappless phase.

\vspace{10pt}
\section{Conclusions and Discussions}
\label{sec_conclusions}

Our study is motivated by the experimental work in Ref. \cite{Stepanov2019}, in which they observe a speed of sound being apparently twice too low. As a result, the experimental results can not be explained upon considering the polariton condensate alone and prompt the questions of how the existence of the reservoir affecting the superfluidity properties of resonantly driven polariton fluids~\cite{Amelio2020,Amelio20201}. Theoretically, this inspires an interesting question as regards the behavior of superfluidity of non-resonant pumped polariton condensate  in the {\it gapped} regime of the Bogoliubov modes, although realizing a polariton condensate in the {\it gapped} regime remains experimentally challenging.

In summary, we calculated the drag force experienced by an impurity  when it moves in a polariton condensate under the non-resonant pumping in the {\it gapped} phase, i.e. the excitations over the condensate are two dissipative gapped condenste's modes and one purely diffusive reservoir's mode.  We find that the part of the drag force contributed by the reservoir's modes is large for small velocity of the impurity, and it gradually decreases as the velocity further increases.
The amplitude of the drag force grows if we reduce the effective diffusive constant, which can be controlled experimentally.

We also find that the part of the drag force contributed by the dissipative gapped modes is nonzero for velocity of the impurity smaller than that required by the Landau's criterion due to the dissipative nature of the condensates.  When the amplitudes of imaginary parts of the dissipative gapped modes go larger, that is, the condensate is more dissipative, the drag force experienced by the impurity becomes larger.
Furthermore, when the velocity of the impurity is larger than that required by the Landau's criterion, the drag force grows approximately linear in velocity, which is similar to that of the traditional condensate in equilibrium. The total drag force may decrease as the velocity of the impurity increases when the velocity is of intermediate value. We hope this work can contribute to the ongoing experiments of studying the non-equilibrium superfluids in the exciton-polariton condensate under non-resonant pumping.

\vspace{10pt}
\acknowledgments
We thank Alexey Kavokin, Y. Xue, C. Gao, Ying Hu, and Yao-Hui Zhu for stimulating discussions.
This work was supported by the National Natural Science Foundation of China (NSFC) under Grants No. 61405003 and the key projects of the Natural Science Foundation of China (Grant No. 11835011) 

\bibliography{myr}

\end{document}